\documentclass[12pt,a4paper]{article}

\usepackage[british]{babel}
\usepackage[a4paper,top=2cm,bottom=2cm,left=2.5cm,right=2.5cm,marginparwidth=1.75cm]{geometry}

\usepackage{amsmath}
\usepackage{graphicx}
\usepackage[colorlinks=true, allcolors=blue]{hyperref}
\usepackage{hyperref}
\usepackage{orcidlink}
\usepackage[title]{appendix}
\usepackage{mathrsfs}
\usepackage{amsfonts}
\usepackage{booktabs} 
\usepackage{caption}  
\usepackage{threeparttable} 
\usepackage{algorithm}
\usepackage{algorithmicx}
\usepackage{algpseudocode}
\usepackage{listings}
\usepackage{enumitem}
\usepackage{chngcntr}
\usepackage{booktabs}
\usepackage{lipsum}
\usepackage{subcaption}
\usepackage{authblk}
\usepackage[T1]{fontenc}    
\usepackage{csquotes}       
\usepackage{diagbox}
\usepackage{multirow}

\usepackage{cite}

\usepackage{setspace}
\usepackage{titlesec}
\titleformat{\section} 
  {\normalfont\Large\bfseries}{\thesection.}{1em}{}
  
\usepackage{float}   
\usepackage{caption} 


\title{\textbf{Formation of Lattice Vacancies and their Effects on  Lithium-ion Transport in  LiBO$_{2}$ Crystals: Comparative \textit{Ab Initio} Studies}}

\author[1,3,**]{Carson Ziemke}
\author[1,2,*,**]{Ha M. Nguyen}
\author[5,*, **]{Sebastián Amaya-Roncancio}
\author[2]{John Gahl}
\author[1,4]{Yangchuan Xing}
\author[1,2,3]{Thomas W. Heitmann}
\author[1,3,*]{Carlos Wexler}
\affil[1]{\small Materials Sciences and Engineering Institute, University of Missouri, Columbia, MO 65201, USA}
\affil[2]{ University of Missouri Research Reactor, University of Missouri, Columbia, MO 65203, USA}
\affil[3]{Department of Physics and Astronomy, University of Missouri, Columbia, MO 65201, USA}
\affil[4]{Department of Chemical and Biomedical Engineering, University of Missouri, Columbia, MO 65201, USA}
\affil[5]{Natural and Exact Sciences Department, Universidad de la Costa, Barranquilla, Colombia}
\affil[*]{Corresponding authors: \texttt{wexlerc@missouri.edu, samaya3@cuc.edu.co, hn4gq@missouri.edu}}
\affil[**]{Authors with equal contributions}

\begin{document}
\maketitle

\begin{abstract}
The monoclinic (m-LBO) and tetragonal (t-LBO) polymorphs of LiBO${_{2}}$ have significant potential for applications such as solid electrolytes and electrode coatings of lithium-ion batteries. While comparative experimental studies of electron and lithium transport in these polymorphs exist, the role of lattice vacancies on lithium transport remains unclear. In this study, we employed density functional theory (DFT) to investigate the impact of boron and oxygen vacancies on the lattice structure, electronic properties, and lithium migration energy barrier ($E_{\text{m}}$) in m-LBO and t-LBO. Our DFT results reveal that boron and oxygen vacancies affect lithium transport in both the polymorphs, but in different ways. While oxygen vacancies lower $E_{\text{m}}$ in m-LBO, they increases  $E_{\text{m}}$ in t-LBO. In contrast, boron vacancies significantly reduce $E_{\text{m}}$ in both m-LBO and t-LBO, leading to enhanced diffusivity and ionic conductivity in both polymorphs. This improvement suggests a potential strategy for improving ionic conductivity in LiBO${_{2}}$ through boron vacancy generation. 

\end{abstract}


\section{Introduction \label{Intro}}

\noindent
Lithium metaborate (LiBO$_{2}$) is an inorganic compound of lithium, boron, and oxygen, which has drawn increasing attention in both pure and applied research \cite{ref1,ref2}. This is due chiefly to its potential applications as an advanced multi-functional material, ranging in use from nuclear reactor materials engineering to nonlinear optics \cite{ref3}, as well as to use in lithium-ion batteries (LIBs). This use for LIBs includes its potential applications as a solid electrolyte \cite{ref4,ref5,ref6,ref7,ref8,ref9} as well as electrode coating materials \cite{ref10,ref11,ref12,ref13,ref14}. Since LiBO$_{2}$ belongs to a large family of boron-contained LIBs materials \cite{ref15,ref16,ref17} that exhibit unique characteristics, establishing, for example, a stable and effective solid electrolyte ininterface (SEI), thermal stability, cost-effectiveness, and environmental friendliness \cite{ref3,ref19,ref20,ref21,ref22,ref23,ref24,ref25}. Therefore, there has been a surge of interest in the mechanistic understanding of defect formation and lithium migration in these materials \cite{ref3,ref19,ref20,ref21,ref22,ref23,ref24,ref25}. Studies intend to shed light on (i) how to improve these materials' ionic conductivity \cite{ref22}, and (ii) how to control radiation-induced modifications to LIBs \cite{ref23,ref24,ref25}. Task (i) focuses on understanding defect chemistry of LIBs materials to engineering the materials for their enhanced lithium-ion diffusion \cite{ref22}. Task (ii) is applied to those LIBs that operate in extreme environments such as in fission/fusion nuclear reactors or deep-space-exploration spacecrafts \cite{ref22, ref23,ref24,ref25} where  neutrons can transmute boron atoms inside LIBs materials via the boron neutron capture reaction (BNCR)  \cite{ref22, ref23,ref24,ref25,ref26,ref27} given as $^{10}_{5}\text{B} + ^{1}_{0}\text{n (0.25 eV)} \rightarrow ^{4}_{2}\text{He (1.47 MeV)} + ^{7}_{3}\text{Li (0.84 MeV)} + \gamma \text{ (0.48 MeV)}$. In this reaction, boron vacancies are created at the original lattice sites of $^{10}_{5}\text{B}$ (comprising ca.\ 20\% of B atoms), while the products of the reaction, which are energetic particles, can trigger atom displacement cascade processes in crystalline solids, resulting in neutron-induced defects. These range from zero-dimensional defects (i.e., point defects in the form of Frenkel pairs) to three-dimensional ones (volume defects in the form of voids or helium bubbles or clusters of defects) \cite{ref3,ref19,ref20,ref21,ref22,ref23,ref24,ref25}. While BNCR is beneficial to certain materials \cite{John2023}, neutron detection techniques \cite{ref26}, and cancer treatment \cite{ref27}, whether it is beneficial or detrimental to boron-containing LIBs working in neutron-based environments is still an open question \cite{ref23,ref24,ref25}. Thus, a deeper insight into the role of boron’s presence and absence (boron vacancy) in LIBs materials in general, and LiBO$_{2}$ in particular, is crucial for both designing and engineering of boron-contained electrode coatings for liquid-electrolyte LIBs or boron-contained solid electrolytes for all-solid LIBs with an optimized content of the $^{10}$B isotope. Moreover, LiBO$_{2}$ is one of the components of SEI formed on the electrode surface of LIBs assembled with borate-based liquid electrolytes that have recently been subject to increasing research interest in the field of liquid electrolytes for LIBs \cite{ref15,ref16,ref17}. Therefore, investigations of lithium transport in LiBO$_{2}$ would also lead to a key fundamental understanding of such a borate-based SEI as well as the overall performance of borate-based LIBs. 

\begin{figure}[!ht]
\centering
\includegraphics[width=1.0\linewidth]{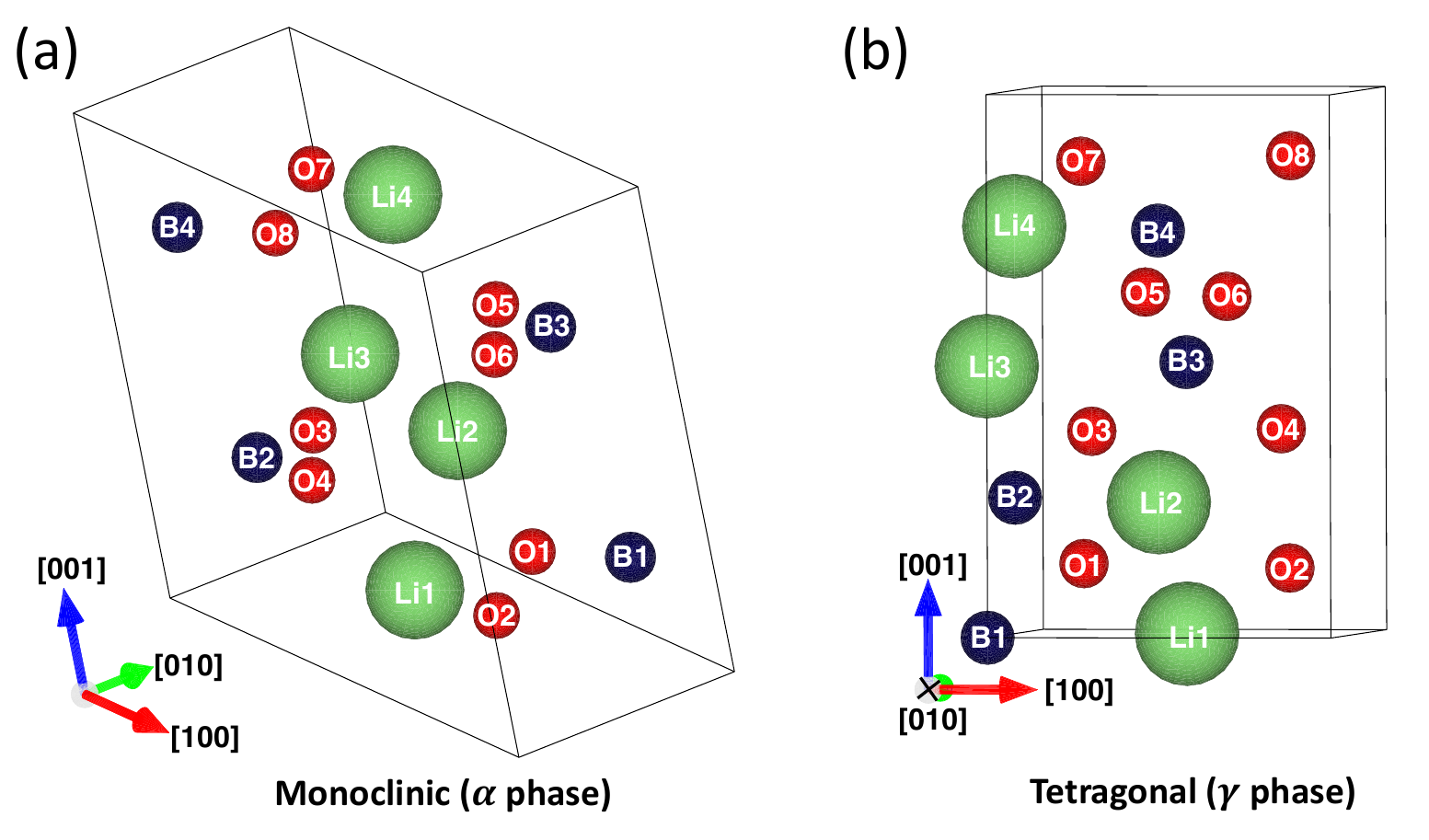}
\caption{\label{Figure1} Unit cells of m-LBO (a) and t-LBO (b) polymorphs of LiBO$_{2}$ crystal.}
\end{figure}

The monoclinic (m-LBO, \text{$\alpha$} phase) and tetragonal (t-LBO, \text{$\gamma$} phase) polymorphs are the most commonly studied polymorphs of the LiBO$_{2}$ crystals (see Figure  \ref{Figure1} and Table \ref{Table1} and the detailed description in Section \ref{Methods}). The m-LBO crystal structure is stable at ambient temperature and pressure. In addition, it was reported that \cite{ref28} baking a hydrated LiBO$_{2}$.8H$_{2}$O powder at 600$^{\circ}$C enables it to crystallize in the ionic m-LBO polymorph. In contrast, the t-LBO polymorph was reported \cite{ref28} to be stable at high pressures. One notable contrast is that while lithium ions in m-LBO form a two-dimensional network, they are arranged in three-dimensional network in t-LBO. To explore how these different configurations influence lithium transport in LiBO$_{2}$ crystals, Hirose \textit{et al.} \cite{ref4} conducted a comprehensive experimental study on high-density polycrystalline samples of these LiBO$_{2}$ polymorphs synthesized under high pressures. The lithium ionic conductivities of the samples were measured using the AC impedance method between 450 and 520 K. They discovered that the intra-grain and total conductivities of t-LBO were of the order 10$^{-6}$ to 10$^{-5}$ S$\cdot$cm$^{-1}$ and 10$^{-6}$ S$\cdot$cm$^{-1}$, respectively, values that were consistently higher than those of m-LBO across the entire measured temperature range.

While theoretical studies have examined the electronic and vibrational properties of the two polymorphs \cite{ref6,ref7} and lithium conductivity in m-LBO \cite{ref5}, the impact of lattice vacancies on lithium transport remains unknown. The present work is aimed at filling this knowledge gap using density functional theory (DFT) calculations. We investigated the effects of boron and oxygen vacancies on the lattice structure, electronic properties, and lithium ion migration energy barrier $E_{\text{m}}$. Our DFT results show that boron and oxygen vacancies affect lithium transport in both polymorphs differently: oxygen vacancies lower $E_{\text{m}}$ in m-LBO but increase it in t-LBO, while boron vacancies significantly reduce $E_{\text{m}}$ in both m-LBO and t-LBO, leading to enhanced diffusivity and ionic conductivity in both polymorphs.

\begin{table*}[!ht]
\caption{Conventional Unit Cells of LiBO$_{2}$ Polymorphs.\label{Table1}}
\tabcolsep=0pt
\begin{threeparttable}
\begin{tabular*}{\textwidth}{@{\extracolsep{\fill}}lcccccc@{\extracolsep{\fill}}}
\toprule
& \multicolumn{3}{c} {\textbf{m-LBO ({\boldmath$\alpha$} phase})}  
& \multicolumn{3}{c}{\textbf{t-LBO ({\boldmath$\gamma$} phase})} \\
& \multicolumn{3}{c} {Monoclinic (P2$_{1}$/c)}   
& \multicolumn{3}{c}{Tetragonal (I$\Bar{4}$2d) } \\
& \multicolumn{3}{c} {$a$ = 5.77 {\AA}, $b$ = 4.46 {\AA}}  
& \multicolumn{3}{c}{$a$ = 4.18 {\AA}, $b$ = 4.18 {\AA}} \\
& \multicolumn{3}{c} {$c$ = 6.37 {\AA} , $\alpha$ = 90$^{\circ}$}    
& \multicolumn{3}{c}{$c$ = 6.53 {\AA}, $\alpha$ = 90$^{\circ}$} \\
& \multicolumn{3}{c} {$\beta$ = 113.56$^{\circ}$, $\gamma$ = 90$^{\circ}$}    
& \multicolumn{3}{c}{$\beta$ = 90$^{\circ}$, $\gamma$ = 90$^{\circ}$} \\
& \multicolumn{3}{c} {$V$=146.98 {\AA}$^{3}$}   
& \multicolumn{3}{c} {$V$=113.86{\AA}$^{3}$}  \\
\cmidrule(lr){2-4}\cmidrule(lr){5-7}
Atom & $x$ ({\AA}) & $y$ ({\AA}) & $z$ ({\AA}) & $x$ ({\AA}) & $y$ ({\AA}) & $z$ ({\AA}) \\
\midrule
 Li1 & 2.879 & 1.256 & 0.885 & 2.090 & 2.090 & 0.000 \\
 Li2 & 1.617 & 3.436 & 2.034 & 2.090 & 0.000 & 1.632 \\
 Li3 & 1.606 & 0.924 & 3.805 & 0.000 & 0.000 & 3.265 \\
 Li4 & 0.344 & 3.104 & 4.954 & 0.000 & 2.090 & 4.897 \\
 B1  & 4.468 & 3.587 & 1.318 & 0.000 & 0.000 & 0.000 \\
 B2  & 0.029 & 1.407 & 1.601 & 0.000 & 2.090 & 1.632 \\
 B3  & 3.195 & 2.952 & 4.238 & 2.090 & 2.090 & 3.265 \\
 B4  &-1.244 & 0.772 & 4.521 & 2.090 & 0.000 & 4.897 \\
 O1  & 3.247 & 3.151 & 1.044 & 0.666 & 3.135 & 0.816 \\
 O2  & 4.667 & 0.620 & 1.396 & 3.514 & 1.045 & 0.816 \\
 O3  &-0.170 & 2.800 & 1.523 & 1.045 & 1.424 & 2.449 \\
 O4  & 1.250 & 0.971 & 1.875 & 3.135 & 2.756 & 2.449 \\
 O5  & 1.974 & 3.389 & 3.963 & 1.424 & 3.135 & 4.081 \\
 O6  & 3.394 & 1.559 & 4.315 & 2.756 & 1.045 & 4.081 \\
 O7  &-1.443 & 3.739 & 4.442 & 1.045 & 0.666 & 5.713 \\
 O8  &-0.023 & 1.209 & 4.795 & 3.135 & 3.513 & 5.713 \\

\bottomrule
\end{tabular*}
\end{threeparttable}
\end{table*}


\section{Computational Methods\label{Methods}}

We utilized density functional theory (DFT) methods, implemented in the Quantum ESPRESSO Package (QE) \cite{ref29}, to investigate the formation of lithium, boron, or oxygen lattice vacancies and lithium migration in both the m-LBO and t-LBO polymorphs. The Perdew-Burke-Ernzerhof (PBE) generalized-gradient approximation (GGA) was employed \cite{ref33}. The effects of lattice vacancies on crystal lattice structure, electronic density of states, and lithium transport were then analyzed from our first-principles calculations.

Our DFT calculations were conducted using plane-wave basis sets of pseudopotentials (UPPs): Li.pbe-sl-rrkjus\_psl.1.0.0.UPF for lithium, B.pbe-n-rrkjus\_psl.1.0.0.UPF for boron, and O.pbe-n-rrkjus\_psl.1.0.0.UPF for oxygen \cite{ref34}. These pseudopotentials have been known for their compatibility with the PBE functionals, ensuring consistency with established theoretical frameworks. Additionally, these pseudopotentials are norm-conserving, accurately representing the core electrons while efficiently capturing the valence electron behavior, crucial for studying lithium diffusion in LiBO$_{2}$ with reduced computational expense and increased reliability \cite{ref35}.

The atomic positions and lattice parameters of conventional unit cells of the polymorphs were obtained from the database of the Materials Project \cite{ref32} 
and are shown in Figure  \ref{Figure1} and Table \ref{Table1}. There are 16 ions per unit cell for each polymorph: 4 lithium cations (labeled Li1 to Li4), 4 boron cations (B1 to B4), and 8 oxygen anions (O1 to O8). On the one hand, the m-LBO polymorph possesses the P2$_{1}$/c space group, where lithium cations (Li$^{+}$) are coordinated with four oxygen anions (O$^{2-}$) in a 4-coordinate geometry, resulting in Li$^{+}$-O$^{2-}$ bond distances ranging from 1.93 to 1.97 {\AA}. Boron ions (B$^{3+}$) exhibit a trigonal planar coordination with three O$^{2-}$ anions, leading to B$^{3+}$-O$^{2-}$ bond distances spanning 1.33 to 1.41 {\AA}. The compound contains two distinct O$^{2-}$ sites: the first site features a distorted trigonal planar arrangement, where one O$^{2-}$ anion connects to one Li$^{+}$ cation and two equivalent B$^{3+}$ cations, while the second site shows O$^{2-}$ anion in a 4-coordinate configuration, bonding with three equivalent Li$^{+}$ cation and one B$^{3+}$ cation. On the other hand, the t-LBO polymorph has the I$\Bar{4}$2d space group. Each of Li$^{+}$ cations is bonded in a 4-coordinate geometry to four equivalent O$^{2-}$ anions. All Li$^{+}$-O$^{2-}$ bond lengths are 1.94 {\AA}. Each of B$^{3+}$ cations is bonded to four equivalent O$^{2-}$ anions to form corner-sharing BO$_{4}$ tetrahedra. All B$^{3+}$-O$^{2-}$ bond lengths are 1.48 {\AA}. Each of O$^{2-}$ anions is bonded in a 4-coordinate geometry to two equivalent Li$^{+}$ cations and two equivalent B$^{3+}$ cations. 

In the current work, we investigated supercells consisting of one unit cell with periodic boundary condition in all three dimensions. The symbol $\square$ represents a vacancy in a crystal structure; a letter in the superscript indicates the type of ion missing, while the subscript number of $\square$ denotes that there is one vacancy per unit cell in defective supercells. As such, Li$_{4}$B$_{4}$O$_{8}$ (16 atoms and no vacancy) is a pristine nondefective supercell, (Li$_{4-1}\square^{\text{Li}}_{1}$)B$_{4}$O$_{8}$ (15 atoms, 1 Li vacancy $\square^{\text{Li}}$) for a Li-vacancy supercell, Li$_{4}$(B$_{4-1}\square^{\text{B}}_{1}$)O$_{8}$ (15 atoms, 1 B vacancy $\square^{\text{B}}$) for a B-vacancy supercell, and Li$_{4}$B$_{4}$(O$_{8-1}\square^{\text{O}}_{1}$) (15 atoms, 1 O vacancy $\square^{\text{O}}$) for O-vacancy supercell. 

Selecting supercells of reduced sizes in DFT calculations entails trade-offs between computational efficiency and system representation. However, such supercells allow for rapid exploration of localized effects of vacancies on lithium transport in LiBO$_{2}$ and facilitate the study of systems with high or even extremely high concentrations of point defects compared to the percolation threshold without losing too much the accuracy of the DFT calculations. To illustrate this point, we provide here an example of a previous DFT calculation of the m-LBO polymorph by  Islam \textit{et al.} \cite{ref5}, who have examined the effect of supercell size on the lithium vacancy formation energy in the m-LBO. They found that the values of the formation energy of lithium vacancy in $1\times 1\times 1$ (\textit{i.e.,} Li$_{4}$B$_{4}$O$_{8}$), $2\times 2\times 2$ (\textit{i.e.,} Li$_{32}$B$_{32}$O$_{64}$), and $3\times 3\times 3$ (\textit{i.e.,} Li$_{108}$B$_{108}$O$_{216}$) supercells are 663 kJ/mol (6.87 eV), 667 kJ/mol (6.91 eV), and 673 kJ/mol (6.98 eV), respectively. From these results, we estimate that our simulation supercells may suffer a $\sim$ 1.5\% accuracy loss of the formation energy vs. a significantly more costly $3\times3\times3$ supercell. This is a reasonable compromise for a fast screening of the many configurations that were sampled in this work as it aims  primarily for a quantitative picture of physics and chemistry of the point defects and their effects on lithium transport. In addition, these supercells serve as valuable benchmark cases for preliminary investigations and provide insights into vacancy-mediated transport phenomena. They could guide subsequent studies with large systems or alternative methodologies ({\textit{e.g.,} classical molecular dynamics modeling of cascade atomic displacement induced with neutron irradiation) to achieve a comprehensive understanding of defect behavior in lithium-containing materials. Moreover, these supercells are advantageous for capturing the physics and chemistry of interactive point defects or their clusters induced by extremely-high-dose radiation such as plasma facing materials in nuclear fusion reactors, shedding light on the material's response under extreme conditions.

The procedure of our DFT calculations is described briefly as follows \cite{ref29}. Firstly, we started our  calculations with the pristine nondefective supercells for each polymorph. The data of the conventional unit cells of both the m-LBO and t-LBO were obtained from the Materials Project (see Table \ref{Table1}). The cutoff energy for the planewave basis functions of 100 Rydberg (1360 eV) and a $5\times5\times5$ Monkhorst–Pack k-point mesh to sample the Brillouin zone of the reciprocal space were used for all supercells.  Plane wave energy cutoff and k-point mesh density were separately tested in our self-consistent calculations and were also converged to give a total energy within 1 meV per supercell. These initialized calculations were then followed by the optimization of the supercells in which the atomic positions and lattice parameters were fully relaxed using the Broyden–Fletcher–Godfarb–Shanno (BFGS) algorithm \cite{ref35} to minimize the total energy until it converges within an accuracy of better than 1 meV per cell and the force convergence criterion is 0.01 eV/{\AA}.

Secondly, each of defective supercells were created by generating a vacancy at a Li, B, or O site in the crystal lattice of the fully-relaxed nondefective supercells. For the m-LBO polymorph, one of Li1, B1, or O1 (see Fig. \ref{Figure1}a and Table \ref{Table1}) was respectively removed to form Li-vacancy, B-vacancy, or O-vacancy supercells. For the t-LBO, to create these defective supercells, Li1, B3, or O5 (see Fig. \ref{Figure1}b and Table \ref{Table1}) was respectively removed. The defective supercells were then fully relaxed to optimize their geometry in the same fashion as those for the nondefective supercells.

Thirdly, the analysis of the effects of the vacancies was performed. The lattice parameters and the volume of each of the supercells were obtained, from which the percentages of the volume change relative to the nondefective supercells were determined. In addition, the formation energy per ion of a supercell of LiBO$_{2}$, $E_{\text{f}}$, was determined as

\begin{equation}
E_\text{f} =\frac{\left(E_\text{tot}-m\times E_\text{Li}-n\times E_\text{B} - k\times E_\text{O} \right)}{m+n+k},
\label{Eq1}
\end{equation}

\noindent
where $E_{\text{tot}}$ is the total energy of the supercell, $m$, $n$, and $k$ are the numbers of Li, B, and O ions in the supercell, $E_{\text{Li}}$, $E_{\text{B}}$, and $E_{\text{O}}$ are respectively the total energy per atom of lithium, of boron, and of oxygen in their solids. In this work, the values $E_{\text{Li}}$ = -202.04 eV (-14.85 Ry), $E_{\text{B}}$ = -85.03 eV (-6.25 Ry), and  $E_{\text{O}}$ = -438.78 eV (-32.25 Ry) were respectively obtained from our DFT calculations of the total energies for a body-centered cubic crystal of lithium metal with the Im$\Bar{3}$m space group, a hexagonal crystal of boron with the P6/mmm space group, and a rhombohedral crystal of oxygen molecules with the R$\Bar{3}$m space group. The setups of these DFT calculations were similar to those for LiBO$_{2}$ described earlier in this section.

The energy of formation of a vacancy X (X = Li, B, or O) in a supercell of LiBO$_{2}$, $E_\text{f}^{\text{X}}$, was calculated as

\begin{equation}
E_\text{f}^{\text{X}} = E_{\text{tot}}^{\text{X vacancy}}-E_{\text{tot}}^{\text{Perfect}}+E_{\text{X}},
\label{Eq2}
\end{equation}

\noindent 
where $E_{\text{tot}}^{\text{X vacancy}}$ and $E_{\text{tot}}^{\text{perfect}}$ are respectively the total energies of the X-vacancy and perfect supercells, $E_{\text{X}}$ is the formation energy of ion X in its solid as afore-mentioned.

To explore the potential migration paths and determine the corresponding migration energy barriers ($E_{\text{m}}$) governing lithium ion diffusion, we utilized the climbing image nudged elastic band (CI-NEB) method\cite{ref36}, implemented within QE \cite{ref29}. The CI-NEB images undergo internal relaxation until the maximum residual force reaches a threshold of less than 0.01 eV/{\AA}, without altering the volume or cell parameters during optimization.

We studied lithium diffusion via the vacancy-mediated mechanism, by which lithium ion randomly hops back and forth between its initial lattice site and a final lithium vacancy site \cite{ref22,ref30}. There are 6 pathways, labeled Path 1 to Path 6 (as shown in Figure \ref{Fig2}), which were searched for the optimal lithium diffusion pathway in the supercells (with and without B or O vacancy) of the m-LBO polymorph. In contrast, 9 pathways, labeled Path 1 to Path 9 (shown in Figure \ref{Fig3}), were examined for the supercells of the t-LBO polymorph. These pathways were also investigated in defective supercells.

\begin{figure}[!ht]
\centering
\includegraphics[width=1.0\linewidth]{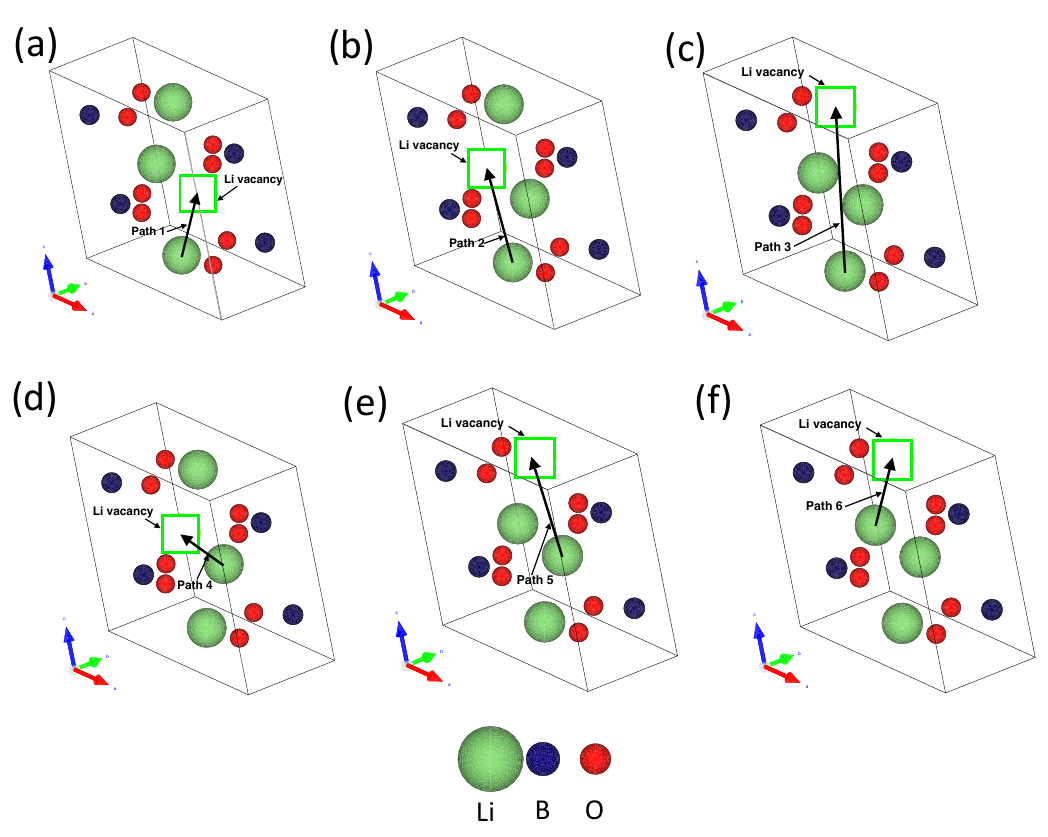}
\caption{\label{Fig2} Li-ion migration pathways investigated in the m-LBO polymorph}
\end{figure}

\begin{figure}[!ht]
\centering
\includegraphics[width=1.0\linewidth]{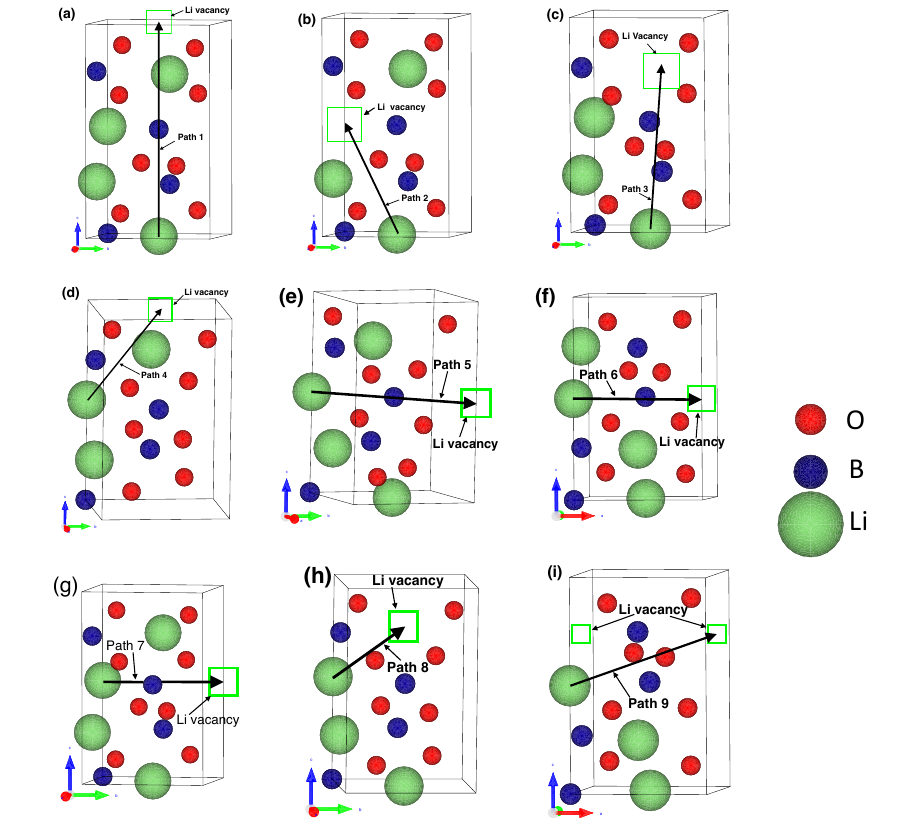}
\caption{\label{Fig3} Li-ion migration pathways investigated in the m-LBO polymorph}
\end{figure}

To compare lithium transport in both m-LBO and t-LBO polymorphs and the effects of O or B vacancy on the transport, the values of the diffusivity of lithium ion along the fastest pathways were determined as

\begin{equation}
D = D_{0}\exp{\left(-\frac{E_{\text{m}}}{k_{\text{B}}T}\right)},
\label{Eq3}
\end{equation}

\noindent
where $E_{\text{m}}$ is the migration energy barrier of the fastest pathway for each of the studied  supercells, $k_{\text{B}}$ is the Boltzmann constant, $T$ is the temperature, $D_{0} \approx a^{2}\nu$, where $a$ is the hopping distance of lithium ion from its initial site to its final vacancy site, and $\nu$ is the lithium hopping frequency. For the sake of simplicity and comparison of the impact of boron or oxygen vacancy on the migration energy barrier and hence the diffusivity, we estimated the same value of $D_{0}$ for all of migration pathways by assigning the values of $a = 5 \text{ \AA}$ and $\nu = 10^{-13} \text{ Hz}$, leading to a reasonable value of $D_{0} = 2.5\times10^{-2}\text{ cm}^2\text{ s}^{-1}$ \cite{ref31}. With this approximation, we would have an idea of the quantitative analysis of how lattice vacancies impact the microscopic local atomic environments of the lattice structure, leading to the modification of the migration energy barrier and hence the modification of transport quantities ($\textit{e.g.,}$ diffusivity, mobility, and conductivity) by several orders of magnitude. Eq. (\ref{Eq3}) enables us to determine the ionic mobility and the ionic conductivity of lithium by using the Einstein–Smoluchowski equation, which relates the diffusivity and mobility of a charged particle (in this case, Li$^{+}$ cation) and is expressed as

\begin{equation}
\mu = \frac{q}{k_{\text{B}}T}D,
\label{Eq4}
\end{equation}

\noindent
where $\mu$ and $q$ are the ionic mobility and the charge of Li$^{+}$ ion, respectively ($q = e =  +1.602\times10^{-19} \text{ C}$). The ionic conductivity of Li$^{+}$ ion is then determined as $\sigma = nq\mu = {q^{2}}nD/({k_{\text{B}}T})$, leading to a direct relation between the conductivity and the migration energy barrier of the fastest lithium migration pathways as

\begin{equation}
\sigma = \frac{q^{2}}{k_{\text{B}}T}nD_{0}\exp{\left(-\frac{E_{\text{m}}}{k_{\text{B}}T}\right)} = \frac{q^{2}}{k_{\text{B}}T}na^{2}\nu\exp{\left(-\frac{E_{\text{m}}}{k_{\text{B}}T}\right)}.
\label{Eq5}
\end{equation}

As stated previously, $n$ is used to denote the concentration of Li$^{+}$ ion and its corresponding vacancy that simultaneously participates with it in the hopping-via-vacancy mechanism of lithium migration. Thus, the vacancy is set to be one per supercell in this work. To compare our estimated values of $\sigma$, we used the same supercell volume of 150 $\text{\AA}^3$, which is the volume of a cubic supercell of $5\text { \AA} \times 5\text { \AA} \times 5\text { \AA}$, for computing the value of $n$. This approach led to the value of $n \approx 8.0\times10^{21}$ cm$^{-3}$. Eq. (\ref{Eq5}) provides a macroscopic-microscopic relationship, which can be measured experimentally using temperature-dependent electrochemical impedance spectroscopy or modeled theoretically using atomistic modeling tools such as the DFT calculations performed in this work. The values of diffusivity, mobility and conductivity are then estimated at $T = 300$ K.

\section{Results and Discussions\label{r&d}}

\subsection{Formation of Lattice Vacancy in LiBO$_{2}$ Crystals}

Islam \textit{et al.} \cite{ref5} have conducted a DFT investigation of lithium vacancies in LiBO$_{2}$ using Perdew Wang (PW91) generalized gradient approximation with projector-augummented waves method (PWGGA-PAWs) implemented in VASP (the Vienna Ab initio Simulation Package) code; they showed that the value of the formation energy is 663 kJ/mol (6.87 eV) for a $1\times1\times1$ supercell of the m-LBO polymorph. They have also attempted to examine both  medium (\textit{i.e.}, $2\times2\times2$) and large (\textit{i.e.}, $3\times3\times3$) supercells. Yet, they only improved the accuracy of the formation energy by respectively 0.6\% and 1.5\% relative to the $1\times1\times1$ supercell. Their work did not explore, however, the formation energies of boron and oxygen vacancies in the m-LBO or t-LBO polymorphs. Therefore, we provide here a comprehensive comparison of the formation energy of each of atomic lattice vacancies in the LiBO$_{2}$ material  by conducting our \textit{ab initio} studies for both of the polymorphs in the same setups of our DFT calculations described in detail in Section \ref{Methods}. Again, to minimize our computational cost, we only employed small (\textit{i.e.}, $1\times1\times1$) supercells. The results are presented in Table \ref{Table2}. It is worth noting from Table \ref{Table2} that, the value of the formation energy of lithium vacancy in the m-LBO polymorph ($\alpha$ phase) is 5.03 eV (485.32 kJ/mol), which is 1.84 eV (177.53 kJ/mol) less than that calculated by Islam \textit{et al.}. This difference is attributed to the different DFT methods and codes used in this work and by Islam \textit{et al.} \cite{ref5} (\textit{i.e.}, our PBEGGA-ultrasoft PPs calculated using the QE code versus their PWGGA-PAWs calculated using the VASP). For the sake of a fair comparison, we have fixed the same setup of our DFT calculations for all of supercells of our interest (See Section \ref{Methods}).

\begin{table}[htbp]
    \centering
    \caption{Formation energy of vacancies in the monoclinic and tetragonal polymorphs of LiBO$_2$.}
    \label{Table2}
    \begin{tabular}{|c|c|c|c|}
        \hline
        \textbf{Polymorph} & \textbf{Vacancy Type} & \textbf{Formation Energy (eV)} \\
        \hline
        Monoclinic & Li Vacancy & $5.03$ \\
                      ($\alpha$ phase)  & B Vacancy & $10.28$ \\
                                      & O Vacancy & $5.74$ \\
        \hline
        Tetragonal & Li Vacancy & $4.64$ \\
                      ($\gamma$ phase)                & B Vacancy & $10.23$ \\
                                      & O Vacancy & $6.49$ \\
        \hline
    \end{tabular}
\end{table}

Table \ref{Table2} presents the values of the formation energy of lithium, boron, or oxygen vacancy in both the m-LBO and t-LBO polymorphs of the LiBO$_{2}$ material. Table \ref{Table2} shows some important points. Firstly, it is energetically more costly to form lithium vacancies in the m-LBO polymorph than in the t-LBO one, which means that the concentration of vacancies at a given temperature is higher in t-LBO than that in m-LBO. In contrast, the formation energies of oxygen vacancies is higher in t-LBO than that in m-LBO. The formation energy of boron vacancies in both polymorphs are the same. Thirdly, the formation energy of vacancies consistently increases from lithium vacancies to oxygen and to boron ones. While the values of the formation energies of lithium and oxygen vacancies in both the polymorphs are considerably high, those of boron vacancies are extremely high, implying that the generation of boron vacancies in the LiBO$_{2}$ material is energetically costly and unfavorable. However, if the LiBO$_{2}$ material is subject to a high dose neutron irradiation, a high concentration of boron vacancies therefore might be possibly achievable in the structure domains of clusters of strongly-interactive vacancies.

\subsection{Effects of Lattice Vacancy on Lattice Structures of LiBO$_{2}$ Crystals} \label{sec3.2}}

This subsection presents the results of our \textit{ab initio} studies of the effects of vacancies in LiBO$_{2}$ polymorphs on the crystal lattice and the energy of formation per atom. The latter provides the understanding of the effects of vacancies on the stability of the crystals.

\begin{table}[htbp]
\centering
\caption{Comparison of theoretical and experimental lattice parameters for monoclinic LiBO\textsubscript{2} and tetragonal LiBO\textsubscript{2}}
\label{Table3}
\begin{tabular}{|c|ccc|ccc|}
\hline
\multirow{1}{*}{\textbf{Parameter}} 
& \multicolumn{3}{c|}{\textbf{Monoclinic LiBO\textsubscript{2} ({\boldmath$\alpha$} phase)}} 
& \multicolumn{3}{c|}{\textbf{Tetragonal LiBO\textsubscript{2} ({\boldmath$\gamma$} phase)}} \\ \cline{2-7} 
                           & Our DFT & Other DFT$^{1}$ & Other Exp$^{2}$ 
                           & Our DFT & Other DFT$^{1}$  & Other Exp$^{2}$ \\ \hline
$a$ (\AA)                  & 5.71    & 5.68 & 5.85                
                           & 4.21    & 4.16 & 4.20            \\ \hline
                           
$b$ (\AA)                  & 4.49    & 4.37 & 4.35             
                           & 4.21    & 4.16 & 4.20            \\ \hline
                           
$c$ (\AA)                  & 6.65    & 6.52 & 6.46             
                           & 6.61    & 6.33 & 6.51            \\ \hline

$\alpha$ ($^{\circ}$)      & 90      & N/A & N/A              
                           & 90      & N/A & N/A            \\ \hline

$\beta$ ($^{\circ}$)       & 110     & N/A & 115.04                
                           & 90      & N/A & N/A            \\ \hline

$\gamma$ ($^{\circ}$)      & 90      & N/A & N/A                
                           & 90      & N/A & N/A            \\ \hline
\end{tabular}
\begin{tablenotes}
\item[1] $^{1}$ Data taken from a theoretical work by Masalaev \textit{et al.} \cite{ref6}. 
\item[2] $^{2}$ Data taken from an experimental work by Hirose \textit{et al.} \cite{ref4}. 
\item[3] N/A means data are not available.
\end{tablenotes}
\end{table}

For the sake of comparison between the experimental and theoretical data of the lattice parameters, Table \ref{Table3} presents the completed data of the lattice parameters of the crystals of both the m-LBO and t-LBO polymorphs theoretically calculated from our current work and the uncompleted data of the parameters reported theoretically by Masalaev \textit{et al.} \cite{ref6} and experimentally by Hirose \textit{et al.} \cite{ref4}. Our data of the lattice parameters of the LiBO$_{2}$ polymorphs were obtained after the supercell of each polymorph was fully relaxed using our PBEGGA-UPP setup in the QE code from the initial inputs shown in Table \ref{Table1}. The data reported by Masalaev \textit{et al.} \cite{ref6} were obtained in their DFT calculations using the CRYSTAL code (CRYSTAL14) with a hybrid B3LYP method, including Becke exchange functional (B3) and the Lee, Yang, and Parr (LYB) correlation method of special points with a $16\times16\times16$ grid and a cutoff energy of 40 Ry. Their DFT calculations were obviously computationally demanding and are expected to better simulate the structural properties of experimental samples. We recall that the experimental data reported by Hirose \textit{et al.} \cite{ref4} were obtained from the Rietveld refinement of their synchrotron powder X-ray diffraction measurements of their LiBO$_{2}$ samples synthesized under high-pressure conditions. It is worth noticing that the relative difference of each of the three lattice parameters, $a$, $b$, and $c$ between our theoretical results and Hirose \textit{et al.}'s experimental values for the m-LBO crystal are $-2.39$\%, $+3.22$\%, and $+2.94$\%, i.e., similar in magnitude to the differences between Masalaev \textit{et al.}'s theoretical and Hirose \textit{et al.}'s experimental values, $-2.91$\%, $+0.46$\%, and $+0.93$\%. 

In contrast, the comparison for the lattice parameters $a$ and $c$ of the t-LBO crystal are as follows ($a = b$ for a tetragonal crystal): $+0.24$\% and $+1.54$\% for our calculation vs. experiment, compared to $-0.95$\% and $-2.76$\% for Masalaev's \textit{et al.} results. More interestingly, our values of $a$ and $c$ are closer to the experimental ones than the Masalaev \textit{et al.}'s ones. This comparison would reinforce that the norm-reserving pseudopotentials chosen in the current work accurately represented the core electrons while efficiently capturing the valence electron behavior, leading to efficient and reliable modeling of LiBO$_{2}$ polymorphs. Overall, the quantitative comparison carried out in this subsection would reveal that the lattice parameters theoretically calculated in the work by Masalaev \textit{et al.} \cite{ref6} and in this work by the authors agreed reasonably well with those reported in the experimental work by Hirose \textit{et al.} \cite{ref4} within a tolerable 5\%-off accuracy.

Table \ref{Table4}  provides the effect of vacancies on the lattice parameters and the volumes of the conventional unit cells in the crystal polymorphs. As one can see on Table \ref{Table4}, after the fully-relaxed optimization of the supercells' geometry in our DFT calculations, some of lattice parameters were modified. As a result, the volumes of the unit cells of the defective supercells were increased or decreased by 1.5\% to 9.8\% relative to those of the corresponding nondefective perfect supercells after their geometry optimization. Specifically, the effects of the lattice vacancies on the crystal structure of each polymorph are separately detailed as follows. For the m-LBO polymorph, the tendency of the effect of each type of vacancies on lattice parameters are consistent: lattice vacancies result in the decreased values of both $a$ and $b$, but increased value of $c$. Nevertheless, the overall impact of vacancies on the volume of the unit cell are different: boron vacancies cause the crystal to expand by $1.5$\% while lithium and oxygen vacancies shrink the crystal by $2.3$\% and $3.6$\%, respectively. Yet, the change of the volume of the unit cell of the m-LBO polymorph is still less then $5$\%.  Notably, the effect of boron and oxygen vacancies on the crystal lattice of t-LBO polymorph is more pronounced, but in opposite direction: nearly 10\% of volume changes in both oxygen-vacancy ($-8.4$\%) and boron-vacancy ($+9.8$\%) unit cells relative to the perfect unit cell, which is quite remarkable. Finally, while the tendency of lithium vacancies to decrease $a$ and $b$ and to increase $c$ in both polymorphs are similar, lithium vacancies shrink the volume ($-2.3$\%) of the m-LBO crystal, but expand that ($+0.3\%$) of the t-LBO one.
 

\begin{table}[!ht]
    \centering
    \caption{Lattice parameters and volumes for nondefective and defective unit cells of the two polymorphs of LiBO$_2$}
    \label{Table4}
    \begin{tabular}{|c|c|c|c|c|c|c|c|}
        \hline
        \textbf{Polymorph} & \textbf{Crystal} & \textbf{a (\AA)} & \textbf{b (\AA)} & \textbf{c (\AA)}  & \textbf{$\beta(^{\circ})$}  & \textbf{V (\AA$^{3}$)} & \textbf{$\Delta V(\%)$} \\
        \hline
         & Perfect & 5.71 & 4.49 & 6.65 & 110.0  & 156.73 & 0.0 \\
        Monoclinic & Li Vacancy & 5.68 & 4.36 & 6.77 & 111.4  & 153.05 & $-2.3$ \\
        ($\alpha$ phase)& B Vacancy & 5.57 & 4.41 & 6.92 & 110.0   & 159.11 & $+1.5$ \\
        & O Vacancy & 5.66 & 4.17 & 6.76 & 108.4  & 151.09 & $-3.6$ \\
        \hline
         & Perfect & 4.21 & 4.21 & 6.61 & 90 & 117.16 & 0.0 \\
        Tetragonal & Li Vacancy & 4.20 & 4.20 & 6.66 & 90 & 117.48 & $+0.3$ \\
        ($\gamma$ phase) & B Vacancy & 4.23 & 4.23 & 7.19 &  90 & 128.65 & $+9.8$ \\
        & O Vacancy & 4.25 & 4.09 & 6.39 & 90  & 107.35 & $-8.4$ \\
        \hline
    \end{tabular}
\end{table}



\begin{table}[!ht]
    \centering
    \caption{Formation energy per ion ($E_{\text{f}}$) in supercells and its absolute change ($\Delta E_{\text{f}}$) relative to perfect nondefective supercells for the two polymorphs of LiBO$_{2}$. $E_{\text{f}}$ is determined using Eq.\ (\ref{Eq1}). $\Delta E_{\text{f}} = E_{\text{f}}^{\text{X vacancy}}-E_{\text{f}}^{\text{Perfect}}$, where X is either Li, B, or O, and $E_{\text{f}}^{\text{perfect}}$ and $E_{\text{f}}^{\text{X vacancy}}$ are energy of formation per ion for nondefective perfect supercell and the X-vacancy counterpart, respectively.}
    \label{Table5}
    \begin{tabular}{|c|c|c|c|}
        \hline
        \textbf{Polymorph} & \textbf{Crystal} & \textbf{$E_{\text{f}}$ (eV/ion)} & \textbf{$\Delta E_{\text{f}}$ (eV/ion)} \\
        \hline
               & Perfect & $-2.45$ & 0.0 \\
         Monoclinic     & Li Vacancy & $-2.28$ & $0.17$ \\
         ($\alpha$ phase)      & B Vacancy & $-1.93$ & $0.52$ \\
               & O Vacancy & $-2.23$ & $0.22$ \\
        \hline
               & Perfect & $-2.45$ & $0.0$ \\
        Tetragonal & Li Vacancy & $-2.30$ & $0.15$ \\
            ($\gamma$ phase)   & B Vacancy & $-1.93$ & $0.52$ \\
               & O Vacancy & $-2.18$ & $0.27$ \\
        \hline
    \end{tabular}

\end{table}

Table \ref{Table5} lists the values of $E_{\text{f}}$ for all of perfect and defective supercells of our interest. These values were determined using Eq.\ (\ref{Eq1}). One can see from Table \ref{Table5} that $E_{\text{f}}$ is the same for the nondefective perfect crystals of both polymorphs. However, when a specific ion (either Li$^{+}$, B$^{3+}$, or O$^{2-}$) is removed from the perfect crystals, $E_{\text{f}}$ increases (\textit{i.e.}, less negativity in its value), which suggests that any lattice vacancy likely results in the reduced energetic stability of LiBO$_{2}$ crystals. In addition, we considered an additional parameter, $\Delta E_{\text{f}}$, which represents the difference in the energies of formation per ion between the defective supercell and its perfect counterpart: $\Delta E_{\text{f}} = E_{\text{f}}^{\text{X vacancy}}-E_{\text{f}}^{\text{Perfect}}$, where X is either Li$^{+}$, B$^{3+}$, or O$^{2-}$, and $E_{\text{f}}^{\text{perfect}}$ and $E_{\text{f}}^{\text{X vacancy}}$ are the energy of formation per ion for the perfect  supercell and defective one, respectively. The higher the value of $\Delta E_{\text{f}}$ is the less the energetic stability of the defective crystal formation relative to its perfect one. Interestingly, the tendency of the destabilization of crystal formation is energetically similar in both the m-LBO and t-LBO polymorphs. Specifically, the degree of destabilization caused by lattice vacancy follows a distinct order in both the polymorphs: $\Delta E_{\text{f}}$ (Li$^{+}$) < $\Delta E_{\text{f}}$ (O$^{2-}$) < $\Delta E_{\text{f}}$ (B$^{3+}$). This trend suggests that the removal of boron ions has the most pronounced destabilizing effect on the LiBO$_{2}$ crystals, followed by oxygen, and lithium has the least destabilizing impact. This trend is consistent with the trend of formation energy of vacancies shown in Table \ref{Table3} and discussed previously. Putting together, the findings of our \textit{ab initio} studies underscore the nuanced and structure-dependent response of LiBO$_{2}$ crystals to specific lattice vacancies, shedding more light on the engineering of defect chemistry to improve the ionic conductivity of LiBO$_{2}$ materials.

\subsection{Effects of Lattice Vacancies on Electronic Insulation and Ionic Conduction of LiBO$_{2}$ Crystals \label{r&dlattice}}

Since LiBO$_{2}$ has been the subject of both experimental and theoretical studies for its use as a solid electrolyte or an electrode coating of Li-ion batteries, insights into how LiBO$_{2}$ material might be engineered, from a practical viewpoint,  to be both a fast Li-ion conductor and a good electron insulator in the same material \cite{ref31}. From a viewpoint of basic research, a better understanding of the impacts of lattice vacancies on the LiBO$_{2}$ material would be essential for the optimization of the materials. This subsection discusses our quantitative analysis of the electrical conduction of LiBO$_{2}$ polymorphs based on the data obtained from our DFT calculations. 

\begin{figure}[!ht]
\centering
\includegraphics[width=1.0\linewidth]{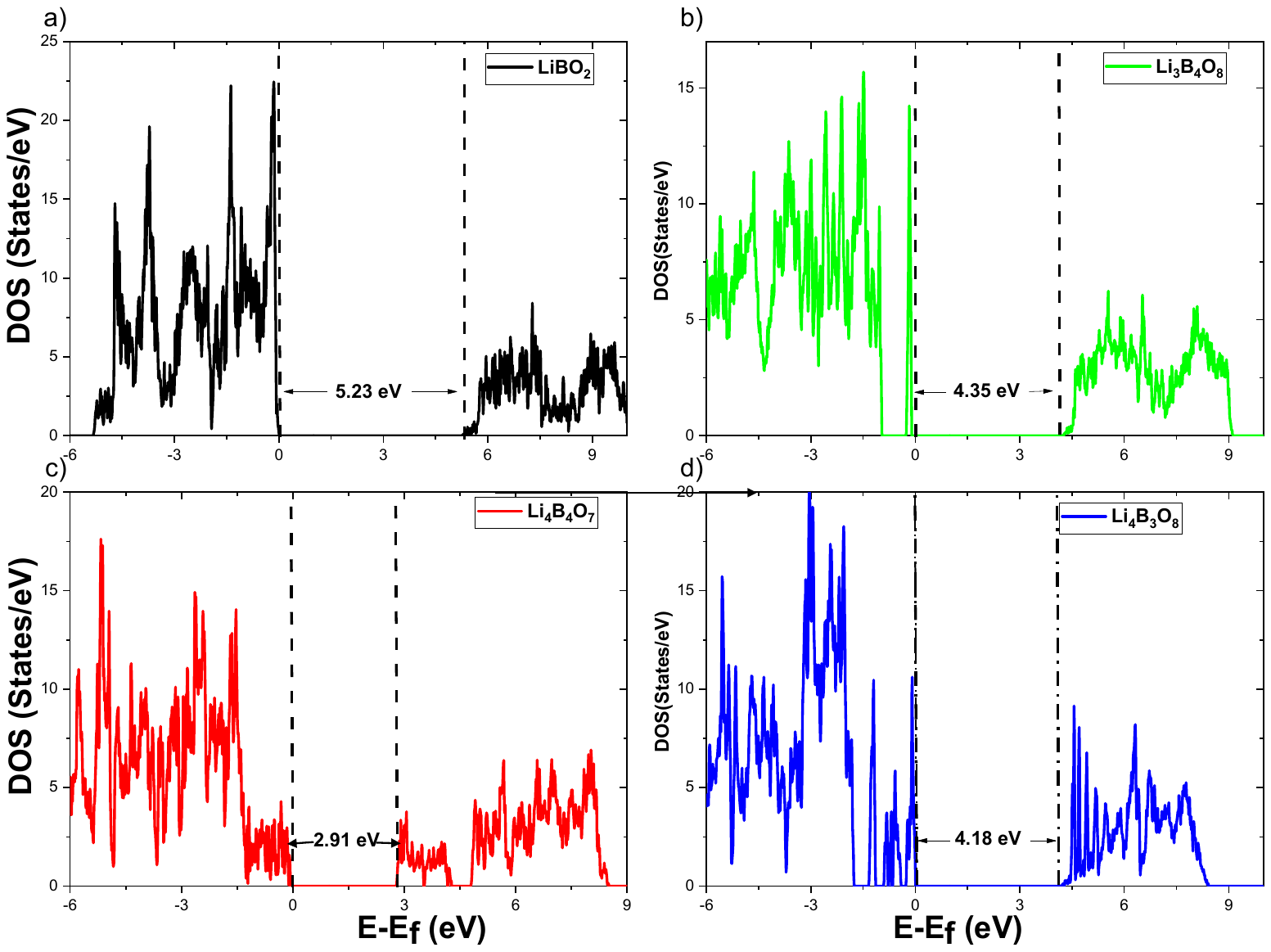}
\caption{\label{Figure4} Electronic Density of states of the m-LBO polymorph. a) Perfect, b) Li vacancy, c) O vacancy, and d) B vacancy.}
\end{figure}

\begin{figure}[!ht]
\centering
\includegraphics[width=1.0\linewidth]{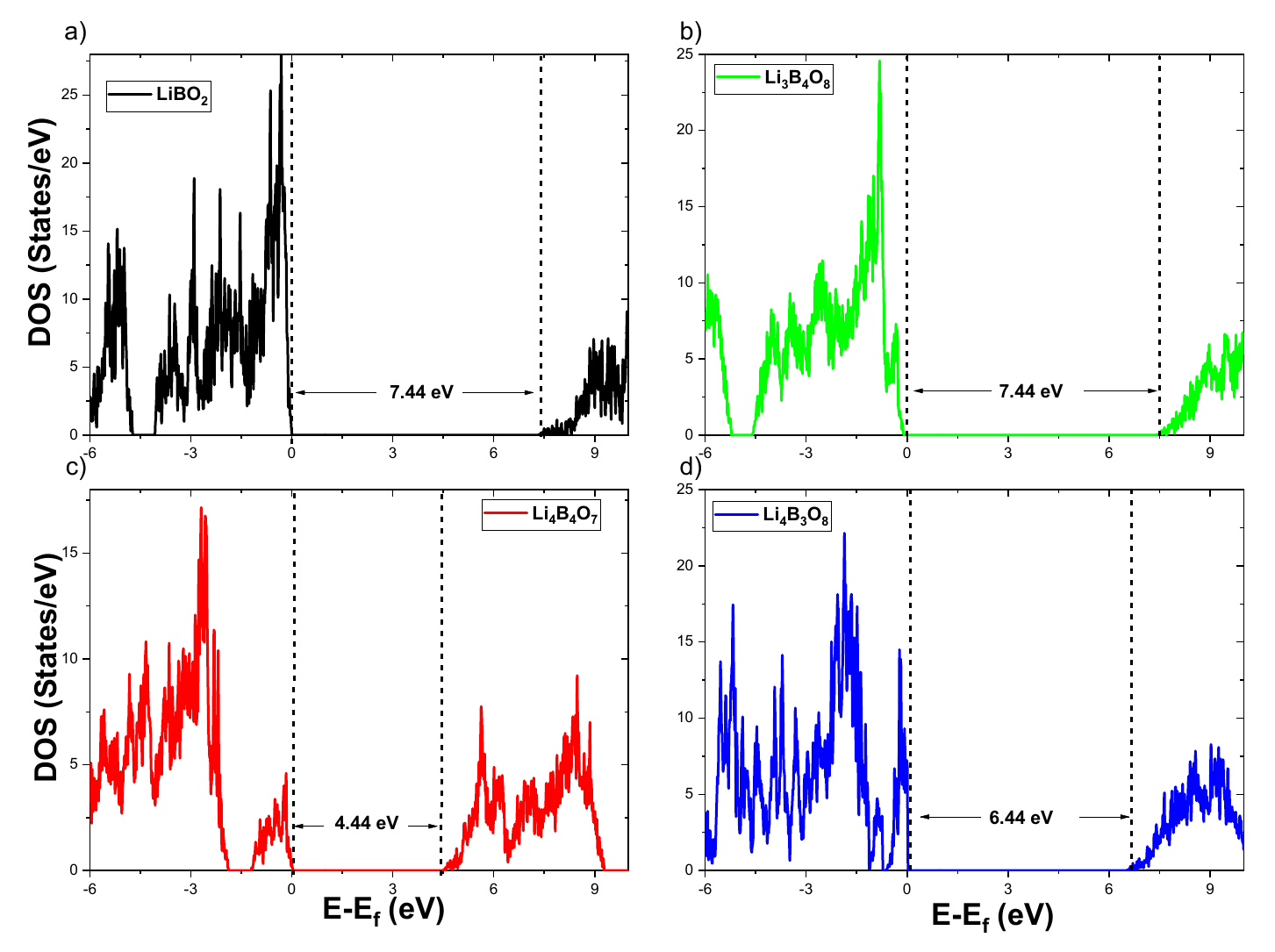}
\caption{\label{Figure5} Electronic density of states of the t-LBO polymorph. a) Perfect, b) Li vacancy, c) O vacancy, and d) B vacancy.}
\end{figure}

\subsubsection{Band Gap and Electronic Insulation}

While experimental data of the band gaps $E_{\text{g}}$ of both LiBO$_{2}$ polymorphs considered here have not been found in the literature, recent DFT calculations (using a hybrid B3LYP method implemented in the CRYSTAL code) for the band structures of the polymorphs by Basalaev \textit{et al.} \cite{ref6} reported that  $E_{\text{g}} = 7.6\text{ eV}$ and $10.4\text{ eV}$ were obtained for m-LBO and t-LBO, respectively. Note that another work of the same authors \cite{ref5}, reported that a PWGGA-PAW method implemented in VASP code yielded a lower value of $E_{\text{g}}= 5.74\text{ eV}$ for the m-LBO polymorph, and that the defect energy levels of the lithium vacancy occurred in the band gap and located adjacent to the top of the valence band, implying that the vacancy is capable of acting as an acceptor of electrons to become negatively charged and the lithium-vacancy m-LBO crystal in this case behaves like a (wide-band-gap) p-type semiconductor. However, there have not been any DFT calculations of electronic band structures and density of states (DOS) published for other types of lattice vacancies (\textit{i.e.}, oxygen and boron vacancies) in the m-LBO polymorph and all types of lattice vacancies (\textit{i.e.}, lithium, oxygen, and boron vacancies) in the t-LBO polymorph. In this subsection, we aim at filling this knowledge gap by presenting the effects of all types of vacancies on DOS, from which one might have deeper insights into the practice of how the LiBO$_{2}$ materials, which function as electronic insulators, can be appropriately designed and engineered to meet technical requirements of desired solid electrolytes or electrode coatings of Li-ion batteries.

 The results of our DOS calculations for the perfect (\textit{i.e.,} nondefective) and defective crystals of LiBO$_{2}$ material are presented in Figures \ref{Figure4} (m-LBO polymorph) and \ref{Figure5} (t-LBO polymorph). Furthermore, panels \ref{Figure4}a and \ref{Figure5}a show that the values of the band gap for the respective perfect crystals are $E_{\text{g}} = 5.23\text{ eV}$ (m-LBO) and $E_{\text{g}} = 7.44\text{ eV}$ (t-LBO).   While these values are less than those reported by  Basalaev \textit{et al.} \cite{ref6} (discussed in the previous paragraph), our results also show that the perfect t-LBO polymorph possesses a wider band gap than that of its perfect m-LBO counterpart. The smaller values of $E_{\text{g}}$ of our DFT calculations are due to the nature of the GGA method which usually underestimates the band gap. Basalaev \textit{et al.} \cite{ref6} used a hybrid B3LYP method which is better at estimating the band gap, but computationally much more costly (which becomes prohibitive given the number of configurations probed in this study).

 From Figures \ref{Figure4} and \ref{Figure5}, the effect of each of the vacancies on DOS in the m-LBO and t-LBO polymorphs is different. Compared to the perfect m-LBO crystal, the DOS of the Li-vacancy m-LBO crystal (panel \ref{Figure4}b) shows the energy levels of lithium defects in the band gap, which is just above and well separated (about $0.8\text{ eV}$) from the top of the valence band, which behave like acceptor levels. These levels are localized to manifest a sharp peak (rather than a continuous energy band) centered at the acceptor level which is 4.35 eV from the botom of the conduction band. The appearance of the defect levels  makes the overall band gap effectively narrower (i.e., 4.35 eV versus 5.23 eV). Therefore, the capability of insulating the electron conduction is weaker for the lithium-vacancy crystal of the m-LBO polymorph than that of its perfect crystal counterpart. In contrast, for the t-LBO polymorph lithium vacancies do not change the value of the band gap (panel \ref{Figure5}b). The defect levels of the lithium vacancies merge together with the top of the valence band, resulting in an effective valence band whose topmost level is still 7.44 eV apart from the bottom of the conduction band. As a result, the band gap is effectively the same as that of the perfect crystal (\textit{i.e}, $E_{\text{g}}= 7.44\text{ eV}$), suggesting that the electronic insulation capability of the t-LBO polymorph might not be deteriorated with the presence of lithium vacancies. 

 The DOS of the boron-vacancy crystals in both polymorphs (panels \ref{Figure4}d and \ref{Figure5}d) manifests a similar scenario to that of lithium-vacancy crystals, where boron defect energies levels occur in the band gap, close to and separate from the top of the valence band (panel \ref{Figure4}d) for the m-LBO polymorph. These defect levels merge altogether with the valence band (panel \ref{Figure5}d) in the t-LBO polymorph. Unlike the lithium-vacancy crystals, the defect energy levels of boron vacancies are broadened to exhibit an overlapping band of the localized levels spanning over approximately a width of 1.0 eV for both polymorphs. For the m-LBO polymorph (panel \ref{Figure4}d) the lowest level of the boron's defect energy levels is nearly 0.4 eV above the top of the valence band while the highest level is about 4.18 eV below the bottom of the conduction band. Consequently, the boron vacancies behave like electron acceptors, leading to a reduced effective band gap of $E_{\text{g}} = 4.18\text{ eV}$, lower than that of the Li-vacancy crystal ($E_{\text{g}} = 4.35\text{ eV}$) as seen in panel \ref{Figure4}b. This means that boron vacancies reduce the capability of electron insulation of the m-LBO polymorph more than lithium vacancies. In the t-LBO polymorph, the band of the defect levels of boron vacancies merge with the top of the valence band and the band gap gets narrower ($E_{\text{g}} = 6.44\text{ eV}$ compared to $E_{\text{g}} = 7.44\text{ eV}$). Similar to the m-LBO polymorph, boron vacancies also tend to degrade the capability of electronic insulation of the t-LBO polymorph.

The effects of oxygen vacancies on DOS (panels \ref{Figure4}c and \ref{Figure5}c) are  fairly different in the two polymorphs. On the one hand, in the m-LBO polymorph (panel \ref{Figure4}c), the defect energy levels of oxygen vacancies are separated into two distinct continuous bands in the band gap with an effective energy gap of $E_{\text{g}} = 2.91\text{ eV}$, which is much narrower than that ($E_{\text{g}} = 5.23\text{ eV}$) of the perfect crystal. The low-energy-level band is approximately 1.2 eV in width and merges with the top of the valence band to create an effective valence band. The high-energy-level band spans as wide as 1.5 eV, and the top of which is located 0.8 eV below the bottom of the conduction band, suggesting that the oxygen-vacancy m-LBO crystal likely behaves like a n-type semiconductor in which the oxygen vacancies could be electron donors that when ionized contribute electrons to the conduction band. In terms of the reduction of the band gap, oxygen vacancies tend to degrade the electronic insulation capability of the m-LBO polymorph even more than boron vacancies. On the other hand, in the t-LBO polymorph (panel \ref{Figure5}c), the defect energy levels of oxygen vacancies exhibit only one continuous band, which is $0.7\text{ eV}$ separated from the top of the valence band, suggesting that oxygen vacancies behave like electron acceptors in a p-type semiconductor. The defect energy band spans in the band gap toward high energies and makes the band gap shrunk to have a narrower band gap of $E_{\text{g}} = 4.44\text{ eV}$, also giving rise to a worsened electronic insulation capability of the t-LBO polymorph.

\subsubsection{Lithium-ion Transport \label{r&dlattice}}

This subsection presents our \textit{ab initio} studies on the effect of lattice vacancies on lithium transport in the two polymorphs of the LiBO$_{2}$ material. These studies constitute the primary focus of our current work. As aforementioned in Section \ref{Methods} and recalled here, a certain number of migration pathways (see Figure \ref{Fig2} and Figure \ref{Fig3}) of lithium ion in each of the studied supercells (either a nondefective or a defective crystal of both the polymorphs of the  LiBO$_{2}$ material) were investigated in order to search for the fastest pathway that possesses the smallest value of the migration energy barrier, $E_{\text{m}}$. We examined the vacancy-mediated diffusion mechanism, where a lithium ion migrates from its initial position to a neighboring lithium lattice vacancy. Simultaneously, the vacancy moves in the opposite direction by swapping its site with the lithium ion. The lithium ion may occupy intermediate sites of varying stability, such as interstitial sites positioned midway during its migration towards the final lithium vacancy site. These intermediate sites will feature mostly higher energies compared to the initial site resulting in significant energy barriers.  The overall rate of the forward hopping for each specific pathway may be estimated by the Arrhenius-type exponential factor $\exp[-{E_{\text{m}}^{\text{f}}}/{k_{\text{B}}T}]$, as depicted in Eq. (\ref{Eq3}) for diffusivity, where $E_{\text{m}}^{\text{f}}$ denotes the highest energy barrier encountered during the forward jumps, defined in this study as the forward migration energy barrier. However, the backward hopping of the associated lithium vacancy partner may encounter the same or different energy barriers, the highest of which is denoted as $E_{\text{m}}^{\text{b}}$ and defined as the backward migration energy barrier.  Only the fastest pathways for both the polymorphs of LiBO$_{2}$ material are presented in this subsection. The other pathways are described in detail in the Appendix.

\begin{figure}[!ht]
\centering
\includegraphics[width=1.0\linewidth]{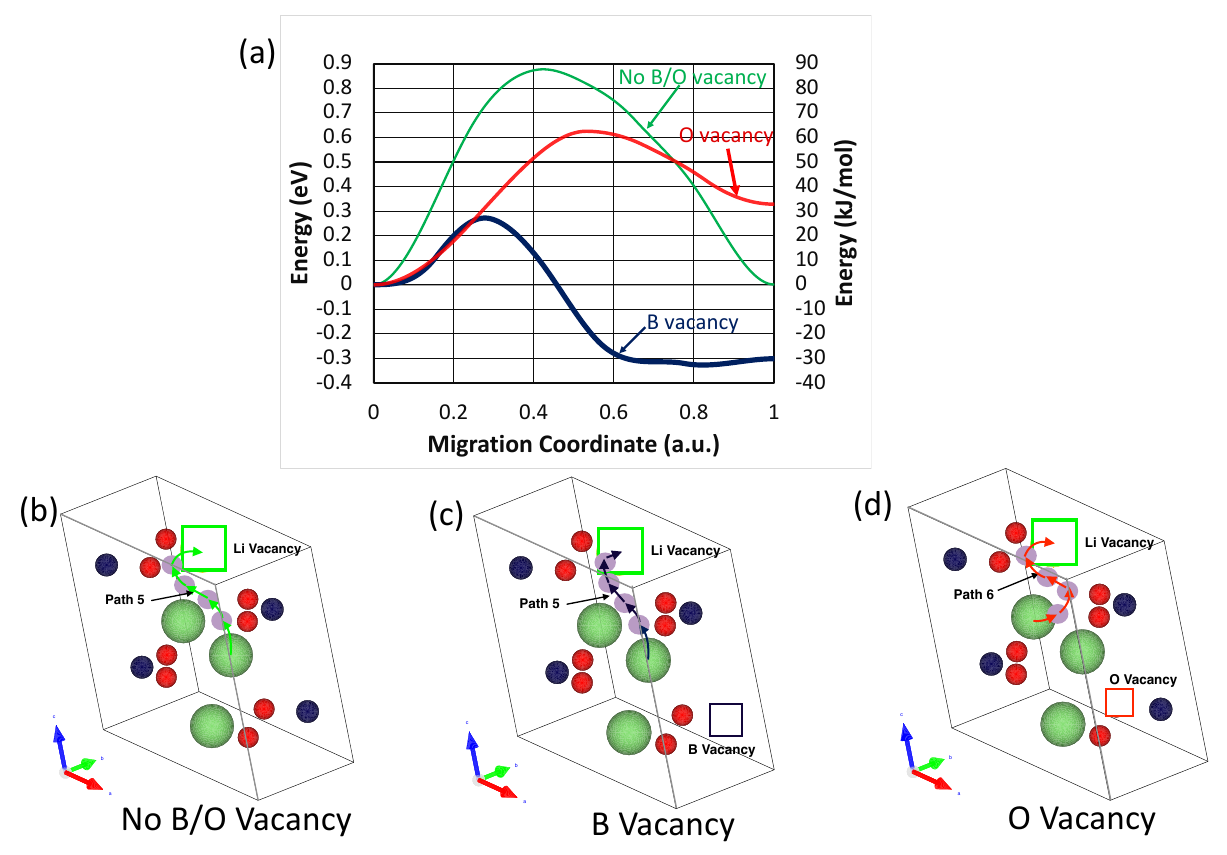}
\caption{\label{Figure6} Fastest Li-ion migration pathways in the m-LBO polymorph.}
\end{figure}
For the six migration pathways under investigation in the m-LBO polymorph, one can see in Figure \ref{Figure6}a that among the three fastest migration pathways, the one in boron-vacancy crystal is of the lowest migration energy barrier. It is suggested from Figure \ref{Figure6}a that the local structure around boron vacancies could facilitate the lithium ion migration in the m-LBO polymorph of LiBO$_{2}$ more efficiently than that of oxygen vacancies, as indicated by the value of $E_{\text{m}}^{\text{f}}$: $E_{\text{m}}^{\text{f}} = 0.88 \text{ eV}$ for the crystal with neither oxygen nor boron vacancies, $E_{\text{m}}^{\text{f}} = 0.27 \text{ eV}$ for the crystal with boron vacancies, and $E_{\text{m}}^{\text{f}} = 0.63 \text{ eV}$ for the crystal with oxygen vacancies. 

Atomic visualizations of the fastest migration pathways are shown in Figures \ref{Figure6}b, \ref{Figure6}c, and \ref{Figure6}d, from which one can note two remarkable observations: (1) all of the fastest migration pathways are not of straight lines, rather curved as S-shapes, and (2) while Path 5 is the fastest migration pathway for both the no B/O vacancy and B-vacancy crystals Path 6 is the fastest pathway for the O-vacancy crystal.

\begin{figure}[!ht]
\centering
\includegraphics[width=1.0\linewidth]{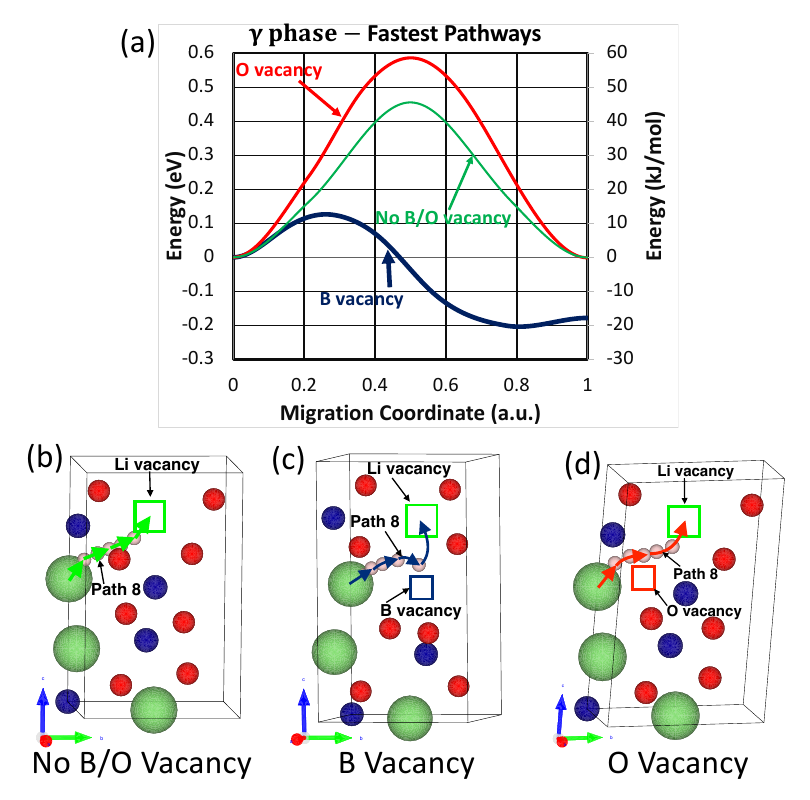}
\caption{\label{Figure7} Fastest Li-ion migration pathways in the t-LBO polymorphs.}
\end{figure}

In comparison, of the nine investigated migration pathways in the t-LBO polymorph, Path 8 becomes the fastest migration pathway in the crystals with and without the presence of boron or oxygen vacancies. As shown in Figure \ref{Figure7}a, while oxygen vacancies increase the energy barrier of Path 8 from $E_{\text{m}}^{\text{f}} = 0.46 \text{ eV}$ to $E_{\text{m}}^{\text{f}} = 0.59 \text{ eV}$, boron vacancies substantially reduce the energy barrier from $E_{\text{m}}^{\text{f}} = 0.46 \text{ eV}$ to $E_{\text{m}}^{\text{f}} = 0.13 \text{ eV}$. The atomistic visualization of these fastest migration pathways are shown in Figures \ref{Figure7}b-c. From these considerations, it is apparent that oxygen vacancies inhibits the migration of lithium ions whereas boron vacancies facilitate it.

\begin{figure}[!ht]
{Table 6. Migration barrier and corresponding diffusivity, mobility, and conductivity of Li-ion transport in LiBO 2 crystals at room temperature estimated for the fastest migration pathways by using density functional theory modeling.}

\includegraphics[width=1.0\linewidth]{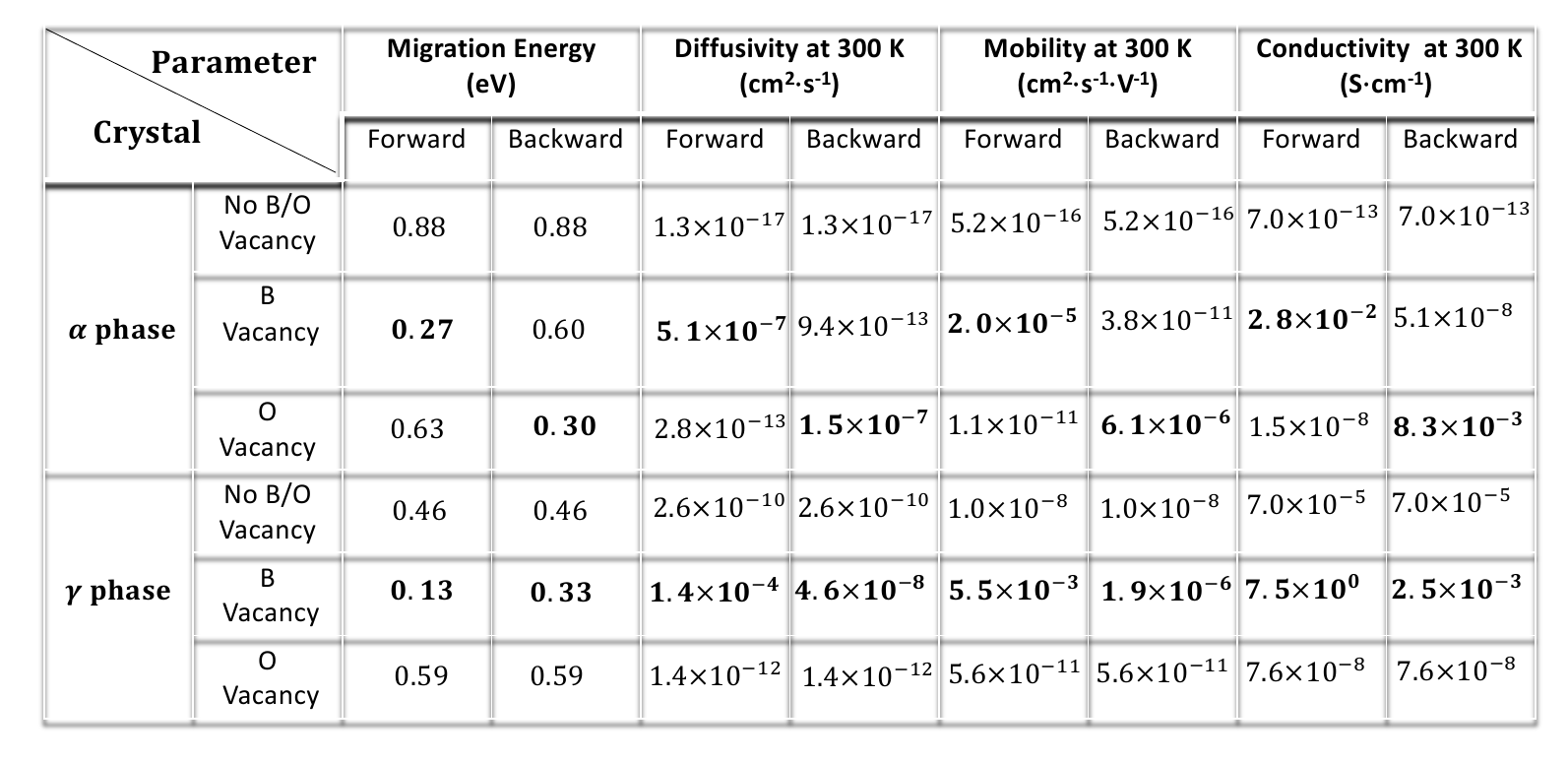}
\end{figure}

The values of $E_{\text{m}}^{\text{f}}$ and $E_{\text{m}}^{\text{b}}$ for the fastest migration pathways of lithium ions in each of the investigated crystals are summarized in Table 6. In addition, the diffusivity, mobility, and conductivity of lithium ions associated with these values are determined using Eqs. (\ref{Eq3}) to (\ref{Eq5}). The purpose of this investigation is to have a quantitative analysis of the microscopic-macroscopic relationship of the impacts of vacancies on lithium ion transport.  As shown in Table 6, our DFT calculations show that in the crystal without boron and oxygen vacancies, the barriers $E^{\text{f}}_{\text{m}} = E^{\text{b}}_{\text{m}} = 0.88$ eV for the fastest pathway are so high that result in extremely low estimated values of the diffusivity ($D =1.3 \times 10^{-17}$ cm$^{2}$ s$^{-1}$) and ionic conductivity ($\sigma = 7.0 \times 10^{-13}$ S cm$^{-1}$) at 300 K. Both boron and oxygen vacancies lower $E_{\text{m}}$, however, in opposite ways due to their opposite charges: ($E_{\text{m}}^{\text{f}}$ = 0.27 eV and $E_{\text{m}}^{\text{b}}$ = 0.60 eV for boron vacancy and $ E_{\text{m}}^{\text{f}}$ = 0.63 eV and $E_{\text{m}}^{\text{b}}$ = 0.30 eV for oxygen vacancy), leading to an increase of the diffusivity and the ionic conductivity by possibly ten orders of magnitude ($D =5.1 \times 10^{-7}$ cm$^{2}$ s$^{-1}$ and $\sigma = 2.8 \times 10^{-2}$ S cm$^{-1}$ for boron vacancy versus $D =1.5 \times 10^{-7}$ cm$^{2}$ s$^{-1}$ and $\sigma = 8.3 \times 10^{-3}$ S cm$^{-1}$ for oxygen vacancy).  In contrast, compared to m-LBO, the effects of boron and oxygen vacancies on lithium transport in t-LBO are quite different. While oxygen vacancy increases $E_{\text{m}}$ from $E_{\text{m}}^{\text{f}} = E_{\text{m}}^{\text{b}} = 0.46$ eV to $E_{\text{m}}^{\text{f}} = E_{\text{m}}^{\text{b}} = 0.59$ eV and inhibits lithium transport, boron vacancy lowers $E_{\text{m}}$ considerably to $E_{\text{m}}^{\text{f}} = 0.13$ eV and $E_{\text{m}}^{\text{b}} =0.33$ eV, giving rise to a remarkable enhancement of both the diffusivity ($D =1.4 \times 10^{-4}$ cm$^{2}$ s$^{-1}$) and ionic conductivity ($\sigma = 7.5 \times 10^{0}$ S cm$^{-1}$) and facilitating very fast lithium transport.

\section{Conclusions}

 The m-LBO and t-LBO polymorphs of LiBO$_{2}$ (m-LBO and t-LBO) have been extensively studied for their diverse technological applications, including as solid electrolytes, solid electrolyte interphase components, and electrode coatings in lithium-ion batteries. Although experimental comparisons of lithium transport in these polymorphs have been made, focusing on their distinct two-dimensional and three-dimensional lithium networks, the mechanistic understanding of point defects, particularly lattice vacancies, on lithium transport remains incomplete. 

The primary goal of the current work is to investigate the formation of lithium, boron, and oxygen vacancies at concentrations  $\sim 8.0 \times 10^{21}$ cm$^{-3}$ and analyzed the resulting modifications in lattice structure, electronic density of states, and lithium migration energy barriers in both polymorphs, focusing on the impact of lattice vacancies on Li-ion transport. Our findings indicate that boron and oxygen vacancies reduce the band gap ($E_{\text{g}}$) of LiBO$_{2}$ crystals by introducing defect levels within the band gap, rendering the material a degenerate semiconductor with a large band gap. In addition, the formation energy of lattice vacancies increase from lithium vacancies, to oxygen and then boron ones.

Regarding lithium ion transport, our DFT results show that oxygen vacancies decrease the migration energy barrier ($E_{\text{m}}$) in m-LBO but increase it in t-LBO. Conversely, boron vacancies significantly reduce $E_{\text{m}}$ in both m-LBO and t-LBO, enhancing diffusivity and ionic conductivity in both polymorphs. This suggests that generating boron vacancies could be a viable strategy for improving ionic conductivity in LiBO$_{2}$. Further theoretical  calculations with larger systems and experimental validation are required to explore this approach fully.

\section*{Conflicts of Interest} 
The authors declare no conflicts of interest.

\section*{Author Contributions}

The contributions of the author to the current work are as follows:
\vspace{-10pt}
\begin{itemize}
\setlength\itemsep{-10pt}
    \item Conceptualization: C.W., S.A., T.W.H, Y.X, J.G.;
    \item Methodology: S.A, H.M.N., C.Z., C.W;
    \item Manuscript Writing: S.A., C.Z., H.M.N., C.W.;
    \item Equal Contribution of DFT Calculations: H.M.N., C.Z., S.A.;
    \item Manuscript Proofreading and Reviewing: all authors.
\end{itemize}

\section*{Acknowledgments}
This work was funded in part by the National Science Foundation Grant No.\ IIP-2044726 and the University of Missouri Materials Science and Engineering Institute (MUMSEI) Grant No.\ CD002339. The computational resource of the University of Missouri-Columbia is acknowledged.

\section*{Appendix: Investigated Pathways of Li-ion Migration in LiBO$_{2}$ Material}

\begin{figure}[!ht]
\centering
\includegraphics[width=1.0\linewidth]{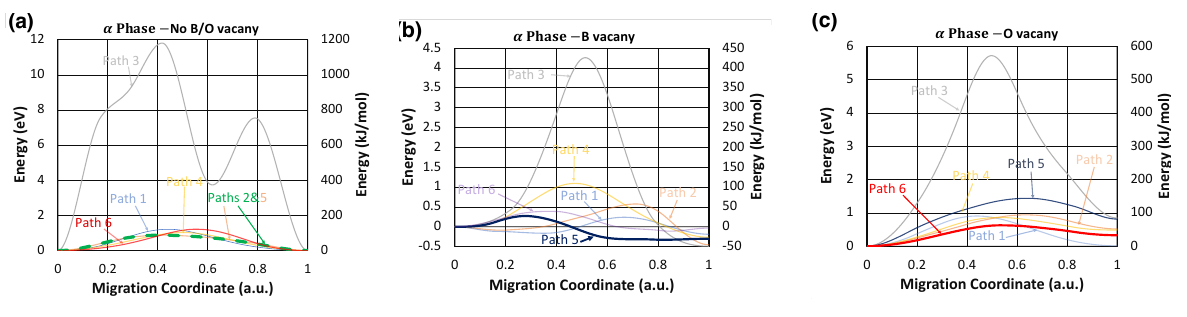}
\caption{\label{Figure8} Energy landscapes of Li-ion migration pathways in the m-LBO polymorph of LiBO$_{2}$ crystal.}
\end{figure}

\begin{figure}[!ht]
\centering
\includegraphics[width=1.0\linewidth]{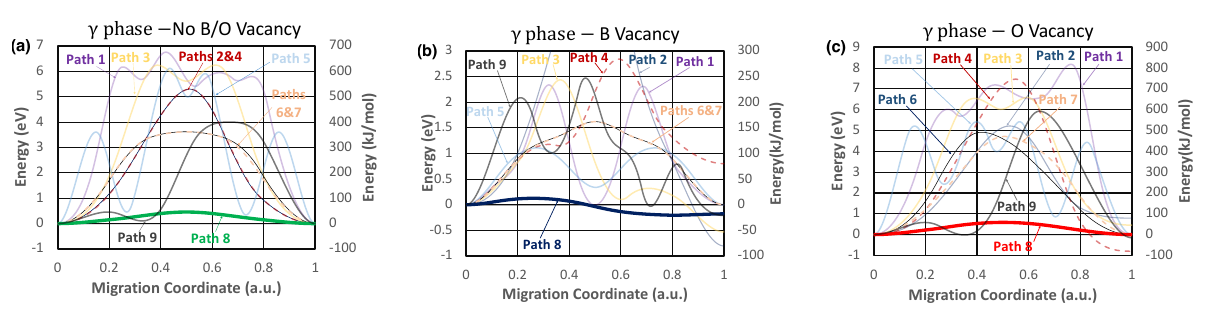}
\caption{\label{Figure9} Energy landscapes of Li-ion migration pathways in t-LBO polymorphs of LiBO$_{2}$ crystal.}
\end{figure}

This appendix presents in detail the energy landscapes of Li-ion migrations pathways other than the fastest pathways in both the polymorphs of the LiBO$_{2}$ discussed in Sec.\ \ref{r&dlattice}. We recall that in order to examine how the local lattice structure at the oxygen or boron vacancies affect the migration of lithium ion, we chose to investigate various pathways near the vacancy sites of our interest (B1 or O1 site for the m-LBO polymorph as labeled in Figure \ref{Figure1}a, and B3 or O5 site for the t-LBO polymorph as in Figure \ref{Figure1}b). Here, 6 migration pathways (Figure \ref{Fig2}) and 9 pathways (Figure \ref{Fig3}) were selected for our search for the optimal migration pathway in each of the examined crystals. The results of the energy landscapes of our CI-NEB calculations for these pathways are depicted in Figures \ref{Figure8} and \ref{Figure9}.

Figure \ref{Figure8} presents the migration energy landscapes for the 6 pathways (Figure \ref{Fig2}) of our interest for crystals without (panel \ref{Figure8}a) and with boron (panel \ref{Figure8}b) or oxygen (panel \ref{Figure8}c) vacancies. These pathways are within the two-dimensional network that includes all of the 4 lithium ions per unit cell of the m-LBO polymorph. As one can see Path 3 is the most prohibitive migration pathway in the m-LBO polymorph under consideration in our current work. For example, for the crystal without boron or oxygen vacancies (panel \ref{Figure8}a), on hopping forward from the lithium site Li1 to the lithium vacant site $\square^{\text{Li4}}$, a lithium ion take two consecutive jumps: the first one is from site Li1 to a less stable interstitial site (about 4 eV higher in energy) requiring the overcoming of an extremely high energy barrier of 12 eV, this intermediate site serves as the "step stone" for the lithium ion to make the second jump to $\square^{\text{Li4}}$ but overcoming a lower barrier of 3.5 eV, thus the forward migration energy of Path 3 is $E_{\text{m}}^{\text{f}} = 12 \text{ eV}$. Along Path 3, the lithium vacancy hops backward from site $\square^{\text{Li4}}$ to site Li1 and encounters two energy barriers, namely, 7.5 eV and 8.0 eV, resulting in a backward migration energy barrier of $E_{\text{m}}^{\text{b}} = 8.0 \text{ eV}$. 

Interestingly, both boron and oxygen vacancies  modify this "two-stop" migration pathway to "one-stop" pathways with substantially reduced energy barriers, approximately three times reduction for boron vacancies (4.5 eV vs. 12 eV) as seen in Figure \ref{Figure8}b and two times by oxygen vacancies (5.6 eV vs. 12 eV) as shown in Figure \ref{Figure8}c. Thus, the vacancies are capable of facilitating the Li-ion migration along Path 3 substantially. Yet, regardless of reduction in the migration energy by both oxygen and boron vacancies, the migration of lithium ion along Path 3 is of extremely low probability and there are better pathways. Due to the migration energies entering into the Arrhenius exponent, ultimately only the lowest migration energy pathways are significant for transport.

It is worth noting in Figure \ref{Figure8}a that the migration energies along other migration pathways are quite lower than those of Path 3. These energies are only a bit more or less than 1 eV. These energies are lowered by the presence of oxygen vacancies (Figure \ref{Figure8}c) and more substantially by the formation of boron vacancies (Figure \ref{Figure8}b).

Figure \ref{Figure9} presents the results of our CI-NEB calculations for 9 lithium migration pathways (see also Figure \ref{Fig3}) in the t-LBO polymorph considered to search for the optimal ones. In contrast to the m-LBO polymorph, lithium ions in the t-LBO crystals create a three-dimensional network. For the crystal without boron or oxygen vacancies (Figure \ref{Figure9}a), there are "nonstop" and "multiple-stop" migration pathways of lithium ions. "nonstop" pathways refer to those along which the lithium ion only overcomes a single energy barrier to hop from the initial lithium site to the final lithium vacancy site. These pathways are Path 2, Path 4, Path 6, Path 7, and Path 8 in Figure \ref{Figure9}a. "Multiple-stop" pathways are those along which one or more "stop(s)" (i.e., interstitial site(s)) serve(s) as the "step stone(s)" for the lithium ion to make its way from its departure site (i.e., the initial lithium site) to its destination site (i.e., the final lithium vacancy site). As shown in Figure \ref{Figure9}a, "multiple-stop" pathways are Path 1, Path 3, Path 5 and Path 9. The migration of lithium ion in the t-LBO polymorph is detailed as follows.

Path 1, which is the most energetically costly pathway, possesses 3 intermediate stops and the destination site with the energies roughly 5.75 eV, 5.25 eV, 5.75 eV, and 0.0 eV higher than the energy of the departure site. The energy barriers that the lithium ion overcomes to take intermediate jumps are approximately 6.15 eV, 1.0 eV, 0.75 eV, 0.2 eV, resulting in $E_{\text{m}}^{\text{f}} = 6.5 \text{ eV}$. The presence of boron vacancies causes Path 1 to be modified remarkably (Figure \ref{Figure9}b): (1) the energetic landscape of Path 1 changes from asymmetric to symmetric; (2) the number of stops is reduced from 3 to 1; and (3) the value of $E_{\text{m}}^{\text{f}}$ is reduced from $E_{\text{m}}^{\text{f}} = 6.5 \text{ eV}$ to $E_{\text{m}}^{\text{f}} = 2.34 \text{ eV}$. In contrast, oxygen vacancies (Figure \ref{Figure9}c) still keep the asymmetric and "multiple-stop" nature and the same value of $E_{\text{m}}^{\text{f}} = 6.5 \text{ eV}$ as Path 1 in the crystal without boron or oxygen vacancies.

Path 3 in the crystal without boron or oxygen vacancies (Figure \ref{Figure9}a) is a symmetric one-stop migration pathway along which lithium ion takes the first jump over an energy barrier of 6.2 eV on hopping forwards from its lithium initial site to an intermediate interstitial site (one stop) 6.66 eV higher in energy that the initial site, and then the second jump over a lower energy barrier of 0.5 eV from the intermediate site to the final lithium vacancy site, resulting in the migration energy barrier $E_{\text{m}}^{\text{f}} = 6.2 \text{ eV}$.  The landscape of Path 3 has not been changed in the crystal with oxygen vacancies (Figure \ref{Figure9}c), its migration barrier is higher ($E_{\text{m}}^{\text{f}} = 6.66 \text{ eV}$) than that of the crystal without boron or oxygen vacancies. However, the symmetry of Path 3 is broken in the boron-vacancy crystal (See Figure \ref{Figure9}b), the lithium ion has to jump over an energy barrier as high as 2.5 eV to an interstitial site of almost the same energy as the initial site, and then takes another jump over a lower barrier of 1.7 eV to its final site which is located at a lower energy (about 0.5 eV lower) than the initial site, leading to $E_{\text{m}}^{\text{f}} = 2.5 \text{ eV}$. Apparently, boron vacancies facilitate lithium transport along Path 3 while oxygen vacancies inhibit it.

Path 5 also possesses a symmetric and multiple-stop (3 intermediate stops) landscape. This landscape is somewhat broken with the presence of oxygen vacancies. The value of $E_{\text{m}}^{\text{f}}$ is a bit lower ($E_{\text{m}}^{\text{f}} = 5.1 \text{ eV}$)  in the oxygen-vacancy crystal (Figure \ref{Figure9}c) than that ($E_{\text{m}}^{\text{f}} = 5.5 \text{ eV}$) in the crystal without boron or oxygen vacancies (Figure \ref{Figure9}a). However, while Path 5 is still symmetric, the number of stops along the pathway is reduced from 3 to 1. Remarkably, boron vacancies reduce the value of $E_{\text{m}}^{\text{f}}$ more than four times, from $E_{\text{m}}^{\text{f}} = 5.5 \text{ eV}$ to $E_{\text{m}}^{\text{f}} = 1.25 \text{ eV}$. Thus, boron vacancies enhance lithium transport along Path 5 more noticeably than oxygen vacancies.

The energy landscape of Path 9 in both the crystal without boron or oxygen vacancies and the crystal with oxygen vacancies looks quite similar (Figures \ref{Figure9}a and \ref{Figure9}c): it is a one-stop migration pathway of lithium ion, starting with a jump from its initial lithium site to an initial site of the same energy level (the only stop) with an energy barrier as low as 0.5 eV, and then another jump from the interstitial site to the final lithium vacant site encountering a much higher energy barrier, the value of which depends on the crystal with or without oxygen vacancies ($E_{\text{m}}^{\text{f}} = 4.0 \text{ eV}$ shown in Figure \ref{Figure9}a for the crystal without boron or oxygen vacancy and $E_{\text{m}}^{\text{f}} = 6.0 \text{ eV}$ shown in Figure \ref{Figure9}c for the crystal with oxygen vacancy). Conversely, Path 9 is modified dramatically with the presence of boron vacancies in the t-LBO. The energetic landscape of the migration pathway becomes an asymmetric two-stop one, energy at any point along the pathway is substantially reduced below 2.5 eV. The description of the migration pathway is specified as follows. On hopping forward along Path 9 in the crystal with boron vacancies, lithium ion takes its first jump from the initial lithium site to an interstitial site (the first stop) at an energy level of 1 eV relative to the initial site and needs to overcome an energy barrier of approximately 2.1 eV. Next, it jumps from the first interstitial site to the second one with the same energy level as the initial site. On this jump, it needs to overcome another energy barrier of 1.5 eV. Finally, it takes the last jump from the second interstitial site to the final lithium vacant site at the energy level that is 0.4 eV lower than the initial site. The energy barrier to overcome for the last jump is approximately 0.75 eV. Overall, the migration energy of lithium ion along Path 9 in the t-LBO crystal with boron vacancies is $E_{\text{m}}^{\text{f}} = 6.0 \text{ eV}$, which is lower than that of the crystal without boron or oxygen vacancies and that of the crystal with oxygen vacancies. Finally, it is worth noting that boron vacancies facilitate lithium transport along Path 9 while the transport is inhibited by oxygen vacancies.

In the t-LBO polymorph, the energy landscapes of Path 2 and Path 4 and those of Path 6 and Path 7, which are nonstop migration pathways of lithium ion, are respectively the same as shown in Figure  \ref{Figure9}a. They are symmetric. However, they are apparently modified with the presence of either boron (Figure \ref{Figure9}b) or oxygen (Figure \ref{Figure9}b) vacancies. In the crystal without boron or oxygen vacancies, Path 2 and Path 4 have the same migration energy barrier of $E_{\text{m}}^{\text{f}} = 6.3 \text{ eV}$ while Path 6 and Path 7 possess a lower barrier of $E_{\text{m}}^{\text{f}} = 5.6 \text{ eV}$. Path 6 and Path 7 still look energetically the same and symmetric in the crystal with boron vacancies even though they are different from those in the crystal without boron or oxygen vacancies. Here boron vacancies lower the value of the migration energy barrier of Path 6 and Path 7 more than twice from $E_{\text{m}}^{\text{f}} = 3.5 \text{ eV}$ (Figure \ref{Figure9}a) to $E_{\text{m}}^{\text{f}} = 1.6 \text{ eV}$ (Figure \ref{Figure9}b). With the presence of oxygen vacancies, the energetic landscapes of Path 6 and Path 7 are no longer the same and symmetric. The energy barriers in Path 6 and Path 7 are respectively $E_{\text{m}}^{\text{f}} = 5.0 \text{ eV}$ and $E_{\text{m}}^{\text{f}} = 6.0 \text{ eV}$.



\end{document}